\documentclass[aps,showpacs,pra,twocolumn]{revtex4-1}

\usepackage{exscale}
\usepackage{graphicx}
\usepackage{amsmath}
\usepackage{latexsym}
\usepackage{amsfonts}
\usepackage{amssymb}
\usepackage{times}
\usepackage[T1]{fontenc}
\usepackage{lipsum}
\usepackage{amsthm}
\usepackage{fancyhdr,txfonts,bbm}

\usepackage{epsfig,pstricks}
\usepackage{changes}

\newtheorem{theorem}{Theorem}
\newtheorem{lemma}{Lemma}

\newtheorem{df}{Definition}
\newtheorem*{tw}{Theorem}

\newcommand{\ot}{{\,\otimes\,}}
\newcommand{{\Cd}}{{\mathbb{C}^d}}
\newcommand{{\C}}{{\mathbb{C}}}
\newcommand{\id}{{\mathcal{I}}}
\newcommand{\SPAN}{{\mathrm{Span}}}

\begin{document}

\title{Constructive method for detecting the information backflow of \\bijective non-completely-positive-divisible dynamics}

\author{Bogna Bylicka$^{1,2}$, Markus Johansson$^1$, and Antonio Ac\'\i n$^{1,3}$}  
\affiliation{$^1$ICFO-Institut de Ciencies Fotoniques, The Barcelona Institute of Science and Technology,
08860 Castelldefels (Barcelona), Spain\\
$^2$Institute of Physics, Faculty of Physics, Astronomy and Informatics,
Nicolaus Copernicus University, Grudzi\c{a}dzka 5/7, 87-100 Toru\'{n}, Poland\\
$^3$ICREA--Institucio Catalana de Recerca i
Estudis Avan\c{c}ats, Lluis Companys 23, 08010 Barcelona, Spain}

\date{\today}
\begin{abstract}

We investigate the relation between two approaches to the characterisation of quantum Markovianity, divisibility and lack of information backflow. We show that a bijective dynamical map is  completely-positive-divisible if and only if a monotonic non-increase of distinguishability is observed for two equiprobable states of the evolving system and an ancilla. Moreover our proof is constructive: given any such map that is not completely-positive-divisible, we give an explicit construction of two states that, when taken with the same a priori probability, exhibit information back-flow. Finally, while an ancilla is necessary for the equivalence to hold in general, we show that it is always possible to witness the non-Markovianity of bijective maps without using any entanglement between system and ancilla.

\end{abstract}

\maketitle

\section{Introduction}

The dynamics of open quantum systems \cite{book_B&P, book_Weiss, book_R&H} has attracted a lot of attention in recent years. In particular the phenomenon of reservoir memory effects and 
the problem of classifying memoryless dynamics, the so-called Markovian regime, and dynamics exhibiting memory effects, the non-Markovian regime, have been investigated extensively (for extended reviews see \cite{rev_RHP, rev_BLP}). However, to date a unique concept of quantum Markovianity does not exist. One can distinguish two main ideas, {\it{completely-positive-divisibility}}, CP-divisability in what follows, and {\it{information flow}}.

The first idea is based on an analogy with the definition of classical Markovian processes. It provides a mathematical characterisation of a map describing a memoryless evolution as a composition of physical maps. This property is known as CP-divisibility and generalises the semigroup property \cite{GKS, L}. Based on this idea a measure of non-Markovianity has been proposed \cite{RHP}, which assesses the deviation of an evolution from being CP-divisible. 

The second idea is more operational and based on the physical features of the system-reservoir interaction, namely, that the phenomenon of reservoir memory effects may be associated with an information backflow. This observation led to the development of a measure of non-Markovianity that quantifies the amount of information that flows back from the environment to the system in terms of the distinguishability of states \cite{BLP}. Following this idea, different measures of information backflow were considered, based for instance on the quantum Fisher information flow \cite{fisher}, the fidelity \cite{fidelity}, the mutual information \cite{Luo}, channel capacities \cite{capacities}, the geometry of the set of accessible states \cite{geometric}, and the channel distinguishability \cite{Joonwoo_Darek}. One can now associate different definitions of Markovianity with each of these measures respectively.  
CP-divisibility implies Markovianity by all the above listed definitions, but the converse is not true for all dynamics \cite{Pinja, RChK, Fifi}. Moreover, it is known that in general the different  definitions do not coincide in the detection of the Markovian regime of the dynamics, which makes the concept of quantum Markovianity even more elusive.

One way of addressing this issue is to assume that reservoir memory effects are a complex phenomenon and a number of measures describing different properties of the open quantum systems are necessary for its full characterisation \cite{zCarole}. On the other hand it raises the essential question of existence of a generalised definition of information flow that would provide a definition of Markovian dynamics equivalent with the mathematical characterisation through CP-divisibility. Some attempts in this direction have been done in Ref. \cite{RChK}. Recently a measure of information flow in direct correspondence with the CP-divisibility property was given in Ref. \cite{Buscemi&Datta}. However a drawback of this measure is the difficulty in its application, as it is not constructive. 

In this work we prove that for bijective dynamical maps, which includes most physically relevant maps, CP-divisibility and information flow are equivalent: for any such dynamical map that is not CP-divisible, it is possible to identify two quantum states, taken with equal prior probability, whose distinguishability increases during the evolution. 
This direct link is enabled by considering distinguishability of states on an extended Hilbert space consisting of the system of interest of dimension $d$ and an additional ancilla of dimension $d+1$. The result is constructive, i.e., for a given dynamical map that is not CP-divisible we show how to derive states displaying an increase of distinguishability during the evolution.  
Finally, despite the fact that the presence of an ancilla is necessary for the equivalence to hold, we show that entanglement is not needed: information backflow can always be observed using two separable states.

\section{Markovian dynamics - definitions}

We consider a quantum system $S$ living on a finite-dimensional Hilbert space $\mathcal{H}_S$ isomorphic to $\Cd$. The set of bounded linear operators acting on $\mathcal{H}_S$ is denoted by $B(\mathcal{H}_S)$, of which the set of states on this Hilbert space, $S(\mathcal{H}_S)$, is a subset. 

The evolution of the quantum system $S$ from initial time $t=0$ to some fixed time $t$ can be described by a dynamical map, i.e., a linear operator $\Lambda_t:S(\mathcal{H}_S)\rightarrow S(\mathcal{H}_S)$ that is completely positive and trace preserving (CPTP).
 The full dynamics of an open quantum system is then given by a time-parametrised family of dynamical maps, $\Lambda=\{\Lambda_t\}$, with the initial condition $\Lambda_0=\id_d$, where $\id_d$ is an identity map.

Let us now discuss in more detail the main two approaches to characterise the memoryless dynamics of open systems.

{\it{Divisibility. -}} The first approach to Markovianity is based on an analogy with the classical Chapman-Kolmogorov equality, which in the classical case is equivalent to the definition of Markovianity for one-point probabilities. In quantum dynamics it is connected to the notion of divisibility.
\begin{df}
A dynamical map $\Lambda_t$ is called divisible if it can be expressed as a sequence of linear maps
\begin{equation}
	\label{eq-div}
	\Lambda_{t} = V_{t,s}\Lambda_s,
\end{equation}
for all times $0 \leq s \leq t$.
\end{df}
Note that if the dynamical map $\Lambda_s$ is invertible, then the intermediate map $V_{t,s}$ is well defined and can be written as $\Lambda_t \Lambda_s^{-1}$. This, however, does not imply (complete) positivity of the map $V_{t,s}$, as the inverse of a CP map in general is not positive. While assuming invertibility is a restriction of the dynamics, it is satisfied for many of the dynamical maps describing physical evolution in which  the equilibrium state is reached asymptotically rather than in a finite time.

In this approach a quantum dynamics is defined as Markovian if and only if it is divisible into a sequence of dynamical maps, namely when the linear map $V_{t,s}$ in equation (\ref{eq-div}) is CPTP. We call such dynamical maps CP-divisible, 

\begin{df}
	A dynamical map $\Lambda_t$ is CP-divisible if and only if it satisfies the decomposition law 
$\Lambda_{t} = V_{t,s}\Lambda_s,$ where $V_{t,s}$ is a completely positive and trace preserving map for all times $0 \leq s \leq t$.
\end{df}

It is also useful to consider the situation in which the intermediate linear map $V_{t,s}$ in Eq. (\ref{eq-div}) is trace preserving and just positive (but not completely positive). Such dynamics is called P-divisible,
\begin{df}
A dynamical map $\Lambda_t$ is P-divisible if and only if
$\Lambda_{t} = V_{t,s}\Lambda_s,$
with $V_{t,s}$ is positive and trace preserving
map for all times $0 \leq s \leq t$.
\end{df}

{\it{Information flow. -}}
The second approach to define Markovianity, introduced in Ref.~\cite{BLP}, is based on a physical interpretation of memory effects in the open-system dynamics as an information exchange between the system and its environment. During a Markovian evolution, information can only flow from the system to the environment, however in the non-Markovian case a temporal backflow, of information previously lost, can occur from the environment to the system. 

\begin{df}
	Information flow between the system, living in $\mathcal{H}_S$, and its environment, during the evolution described by the dynamical map $\Lambda_t$, is defined as
\begin{equation}
	\sigma(\{\rho_i\}, t):= \frac{d}{dt} I(\{\Lambda_t(\rho_i)\}),
\end{equation}
where $\rho_i$ are states on $\mathcal{H}_S$ and $I(.)$ is some relevant information quantifier.
\end{df}
In this approach, an open quantum system dynamics is called Markovian if and only if the distinguishability of states is monotonically non-increasing in time. This definition was originally proposed in Ref.~\cite{BLP} using as information quantifier $I(.)$ the trace distance, $D(\rho_1, \rho_2)=\frac{1}{2}||\rho_1-\rho_2||_1$, where $||.||_1$ stands for the trace norm. This quantifier has a clear operational meaning, as it gives the minimal error probability of distinguishing states $\rho_1$ and $\rho_2$ with the same a priori probability~\cite{Helstrom}. As mentioned, other variants of this measure have been considered for other information quantifiers.

{\it{Previous results. -}} 
Before proving the announced equivalence between CP-divisibility and information flow for bijective dynamical maps, we review some previous results that are relevant for what follows. A detailed discussion of the connection between our work and these results is given below.

First, note that by probing the distinguishability of quantum states without an ancilla, as proposed in~\cite{BLP}, it is impossible to distinguish between P- and CP-divisibility.
This follows from the following result 
 \begin{tw}\cite{koss2, Ruskai}\label{Kossak}
 	Any trace preserving linear map $\mathcal{E}:B(\mathcal{H}) \rightarrow B(\mathcal{H})$ is positive if and only if for any Hermitian operator $\Delta$ acting on $\mathcal{H}$, $||\mathcal{E}(\Delta)||_1\leq ||\Delta||_1 holds$.
 \end{tw}
Note that, up to normalization, any Hermitian operator can be interpreted as the so-called Helstrom matrix \cite{Helstrom}, $\Delta_p = p \rho_1 - (1-p) \rho_2$, characterising the error probability of discriminating between two states $\rho_1$ or $\rho_2$ with unequal probabilities, $p$ and $1-p$ respectively~\cite{RChK}. This extended notion of distinguishability is more than just a formal refinement, as it increases the sensitivity of the information flow definition, in particular it enables detection of the non-Markovian behaviour of the non-unital parts of the dynamical map. Yet, the previous theorem implies that it does not allow distinguishing between P- and CP-divisiblity. Thus, any attempt to connect CP-divisibility and information flow should consider the distinguishability of states on an extended Hilbert space of system and ancilla of at least the same dimension $d$, $\Cd \ot \Cd$.

The previous theorem is also relevant in the case in which the dynamical map is invertible and the intermediate map $V_{t,s}$ can be defined~\cite{RChK}. Let's assume that this map is not CP for some interval of time $[s,t]$. Then, the theorem guarantees the existence of a Helstrom matrix in $\Cd \ot \Cd$, $\Delta^*$, that \emph{witnesses} it, i.e., $||(\mathcal{I}_d \ot V_{t,s})(\Delta^*)||_1 > ||\Delta^*||_1$. However, this result is not enough to guarantee an operational information backflow in the dynamics. For that, one would need to show that  $\Delta^*$ is of the form $p \rho_1 - (1-p) \rho_2$, where $\rho_1$ and $\rho_2$ are states lying in the image of the previous map, $I(\mathcal{I}_d \ot \Lambda_s)$. The problem whether this is always possible has been overlooked in the literature. In other words, it is conceivable a situation in which the intermediate map $V_{t,s}$ is not CP, but no Helstrom operator $\Delta^*$ detecting it can be constructed from states that are reachable during the evolution. Hence, no information backflow, in terms of state distinguishability, would be observed. In Appendix~\ref{app3} we provide an explicit argument showing that invertibility of the dynamical map is a sufficient requirement to ensure backflow of information for any $V_{t,s}$ that is not CP.

A second result relevant in the present context was derived in~Ref.\cite{Buscemi&Datta}. There, a further generalisation of information backflow was defined in terms of the \emph{guessing probability} of discriminating an ensemble of states $\{\rho_i\}$ acting on $\Cd \ot \Cd$ with prior probabilities $p_i$. It was shown that an evolution is CP-divisible if and only if the guessing probability decreases for any ensemble of states. The applicability of this result however is unclear, as the result is not constructive and, in particular, the size of the ensemble witnessing the non-Markovianity is upper bounded only by $d^4$.

\section{Equivalence between CP-divisibility and non-increasing distinguishability for bijective dynamical maps}

After reviewing the relevant definitions and concepts, we are in position to present our main results. We now show that CP-divisibility and monotonic non-increase of distinguishability for all pairs of states of the system and a $(d+1)-$dimensional ancilla, as measured by the trace distance, are equivalent if the dynamical map is bijective. 

\begin{theorem}\label{theo}
A bijective dynamical map $\Lambda_t$ acting on $\Cd$ is CP-divisible if and only if the evolution does not increase the distinguishability, as measured by the trace distance, between any two initial states $\rho_1$ and $\rho_2$ on $\C^{d+1}\ot \Cd $ with the same priori probability for any two times $s<t$; that is $||\mathcal{I}_{d+1}\otimes{\Lambda_s(\rho_1-\rho_2)}||_1\geq ||\mathcal{I}_{d+1}\otimes{\Lambda_t(\rho_1-\rho_2)}||_1$.
\end{theorem}
Theorem \ref{theo} shows that for bijective dynamical maps CP-divisibility of the evolution is equivalent to monotonic non-increase in distinguishability measured by the trace distance, or error probability, already for two equiprobable states.
The proof of Theorem \ref{theo} is the constructive method to find initial states that witness the information back-flow, which is given in Sect. \ref{ex}.

In relation to Theorem \ref{theo} we can make an additional observation: the increase in distinguishability can always be detected without using any entanglement.

\begin{lemma}\label{obs}
If there exist two initial states $\rho_1$ and $\rho_2$, and two times $s<t$, such that the distinguishability as measured by the trace distance increases; that is $||\mathcal{I}_{d+1}\otimes{\Lambda_s(\rho_1-\rho_2)}||_1<||\mathcal{I}_{d+1}\otimes{\Lambda_t(\rho_1-\rho_2)}||_1$, then there always exist two separable states $\rho_1'$ and $\rho_2'$ 
such that $||\mathcal{I}_{d+1}\otimes{\Lambda_s(\rho_1'-\rho_2')}||_1<||\mathcal{I}_{d+1}\otimes{\Lambda_t(\rho_1'-\rho_2')}||_1$.
\end{lemma}
For the proof of Lemma \ref{obs} see Appendix \ref{sep}.
Note that Lemma \ref{obs} holds also for non-bijective dynamical maps.

\section{Constructive method}
\label{ex}
We now give the constructive method to find initial states that witness the information backflow for any evolution described by bijective dynamical maps which is not CP-divisible. This construction also serves as the proof of Theorem \ref{theo}.

If the evolution is described by bijective dynamical maps we can, for each time $s$ and each possible dynamical map $\Lambda_s$, explicitly construct two initial states such that their distinguishability increases for any subsequent evolution $V_{t,s}$ that is not CP. Part of the construction is inspired by the techniques of Ref.~\cite{piani}. 
For this, we consider two orthogonal subspaces $\mathcal{H}_{A}$ and $\mathcal{H}_{A'}$ of the ancilla Hilbert space $\mathbb{C}^{d+1}$, isomorphic to $\mathbb{C}^d$ and  $\mathbb{C}$, respectively. We then make use of the three following observations.

First, consider the maximally entangled state $\phi^{+}=\sum_{i,j}|b_i\rangle\langle b_j|\otimes{|a_i\rangle\langle a_j|}$ where  $|a_i\rangle$ is an orthonormal basis of $\mathcal{H}_{S}$ and likewise  $|b_i\rangle$ is an orthonormal basis of $\mathcal{H}_{A}$. The operator ${\mathcal{I}_{d}}\otimes V_{t,s}(\phi^+)$ is non-negative if and only if $V_{t,s}$ is CP \cite{choi,jammy}. 

Secondly, consider the state $|d+1\rangle\langle d+1|\otimes \rho_{S}\in S({\mathcal{H}_{A'}\otimes{\mathcal{H}_{S}}})$ where $\rho_{S}$ is any state in $S(\mathcal{H}_{S})$. The states $\phi^+$ and $|d+1\rangle\langle d+1|\otimes \rho_{S}$ are orthogonal since $|d+1\rangle\langle d+1|\in S(\mathcal{H}_{A'})$ is orthogonal to any state in $S(\mathcal{H}_{A})$. Moreover, the map $\mathcal{I}_{d+1}\otimes{V_{t,s}}$ preserves the orthogonality, i.e., $\mathcal{I}_{d+1}\otimes{V_{t,s}}(\phi^+)$ is orthogonal to $\mathcal{I}_{d+1}\otimes{V_{t,s}}(|d+1\rangle\langle d+1|\otimes \rho_{S})$.

Lastly, we consider a state $\sigma$ in the interior of the image $I(\mathcal{I}_{d+1}\otimes{\Lambda_{s}})$ of the map $\mathcal{I}_{d+1}\otimes{\Lambda_s}$. By $\sigma$ being in the interior we mean that there exist an open subset of states $X\subset{I(\mathcal{I}_{d+1}\otimes{\Lambda_{s}})}$ such that $\sigma\in{X}$.
Since $\Lambda_{s}$ is bijective the image $I(\mathcal{I}_{d+1}\otimes{\Lambda_{s}})$ is a subset of the set of states that has the same dimensionality as the full set of states. This means that for any state $\rho$ there is a sufficiently small $p$ such that $(1-p)\sigma+p\rho\in I(\mathcal{I}_{d+1}\otimes{\Lambda_{s}})$.

\begin{figure}
\includegraphics[width=0.5\textwidth]{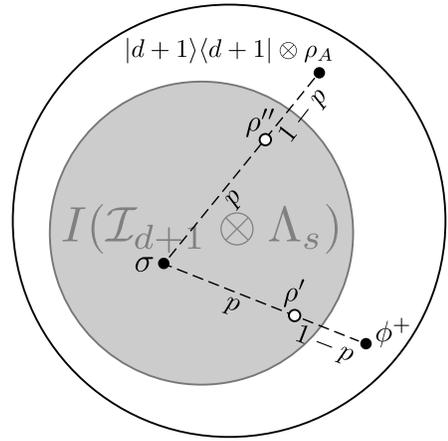}
\caption{The construction of two states $\rho'$ and $\rho''$ inside the image of $I(\mathcal{I}_{d+1}\otimes{\Lambda_{s}})$ (grey area), and detecting that the evolution is not CP-divisible, is obtained by mixing states $\phi^+$ and $|d+1\rangle\langle d+1|\otimes \rho_{S}$ with a state $\sigma$ in the interior of $I(\mathcal{I}_{d+1}\otimes{\Lambda_{s}})$.}
\end{figure}

These three observations allows us to construct two states $\rho'$ and $\rho''$, both evolved from initial states, for which distinguishability increases if $V_{t,s}$ is not CP, in the following way.
Since the state $\sigma$ is in the interior of $I(\mathcal{I}_{d+1}\otimes{\Lambda_{s}})$ there is
some sufficiently small $p$ such that
the two states $\rho'=(1-p)\sigma+p\phi^+$ and $\rho''=(1-p)\sigma+p|d+1\rangle\langle d+1|\otimes \rho_{S}$ are both in $I( \mathcal{I}_{d+1}\otimes\Lambda_s)$.

From the orthogonality of $\phi^+$ and $|d+1\rangle\langle d+1|\otimes \rho_{S}$ it follows that $||\rho'-\rho''||_1=p||\phi^+-|d+1\rangle\langle d+1|\otimes \rho_{S}||_1=2p$. Moreover, $\mathcal{I}_{d+1}\otimes V_{t,s}(\phi^+)$ fails to be positive semidefinite if and only if $V_{t,s}$ is not CP, and if it has a negative eigenvalue it follows that $||\mathcal{I}_{d+1}\otimes{V_{t,s}}(\phi^+)||_1>1$. Since $\mathcal{I}_{d+1}\otimes{V_{t,s}}(\phi^+)$ is orthogonal to $\mathcal{I}_{d+1}\otimes{V_{t,s}}(|d+1\rangle\langle d+1|\otimes \rho_{S})$ it follows that $||\mathcal{I}_{d+1}\otimes{V_{t,s}}(\rho'-\rho'')||_1=p||\mathcal{I}_{d+1}\otimes{V_{t,s}}(\phi^+-|d+1\rangle\langle d+1|\otimes \rho_{S})||_1=p||\mathcal{I}_{d+1}\otimes{V_{t,s}}(\phi^+)||_1+p||\mathcal{I}_{d+1}\otimes{V_{t,s}}(|d+1\rangle\langle d+1|\otimes \rho_{S})||_1$.

Now it is clear that if $V_{t,s}$ is CP, the distinguishability does not change between time $s$ and time $t$, i.e.,
$|| \mathcal{I}_{d+1}\otimes V_{t,s}(\rho'-\rho'')||_1=||\rho'-\rho''||_1=2p$. But if $V_{t,s}$ is not CP, it follows that $||\mathcal{I}_{d+1}\otimes V_{t,s}(\rho'-\rho'')||_1>||\rho'-\rho''||_1=2p$, i.e., the distinguishability increases.
Thus the two states $\rho'$ and $\rho''$ serve as a witness of any evolution in the timestep between $s$ and $t$ that cannot be described by a CP map. Since $\Lambda_s$ is bijective the initial states that evolve into $\rho'$ and $\rho''$ can be constructed as $\Lambda_s^{-1}(\rho')$ and $\Lambda_s^{-1}(\rho'')$. Note that, from Lemma \ref{obs}, it follows that the two initial states can always be constructed as separable states, by choosing the state $\sigma$ from the interior of the image of the separable states under the map $\mathcal{I}_{d+1}\otimes\Lambda_s$ and $p$ small enough.

\section{Non-bijective maps}

We now briefly discuss the case when the dynamical map is not bijective. First of all, note that the constructive method in Sect. \ref{ex} can be applied also to evolutions described by non-bijective dynamical maps for any time $s$ where the map $\Lambda_s$ is bijective, but fails to be applicable for times when the map is not bijective. A non-bijective dynamical map is not necessarily divisible.
However, we show that any evolution that does not increase the distinguishability between any two input states is divisible.

\begin{lemma}\label{lemm2}
If the evolution of the system does not increase the distinguishability between any two initial states $\rho_1$ and $\rho_2$ for any two times $s<t$, as measured by the trace distance; that is if $||\mathcal{I}_{d+1}\otimes{\Lambda_s(\rho_1-\rho_2)}||_1\geq ||\mathcal{I}_{d+1}\otimes{\Lambda_t(\rho_1-\rho_2)}||_1$, the evolution is divisible into a sequence of linear maps.
\end{lemma}
For the proof of Lemma \ref{lemm2} see Appendix \ref{app1}.

If the dynamical map is not bijective at time $s$ there exists two initial states $\rho_1$ and $\rho_2$ such that $\mathcal{I}_{d+1}\otimes\Lambda_s(\rho_1)=\mathcal{I}_{d+1}\otimes\Lambda_s(\rho_2)$.
This means that the dimension of $I(\mathcal{I}_{d+1}\otimes{\Lambda_{s}})$ is lower than the dimension of the full state space.
Furthermore, if the dimension of the image at time $t$ is greater than the dimension of the image at time $s$ there does not exist any map from time $s$ to time $t$ that describes this part of the evolution. Thus the evolution is not even divisible. In this situation Lemma~\ref{lemm2} implies that there always exist two initial states for which the distinguishability increases between time $s$ and time $t$. These two states can be chosen as any two initial states $\rho_1$ and $\rho_2$ such that $\mathcal{I}_{d+1}\otimes\Lambda_s(\rho_1)=\mathcal{I}_{d+1}\otimes\Lambda_s(\rho_2)$ but $\mathcal{I}_{d+1}\otimes\Lambda_t(\rho_1)\neq\mathcal{I}_{d+1}\otimes\Lambda_t(\rho_2)$. Thus for non-bijective dynamical maps where the dimension of the image is not monotonically non-increasing we can always find initial states for which the distinguishability increases at some time.

The case that remains is the non-bijective maps for which the dimension of the image is monotonically non-increasing. Then it is not clear if two states with increasing distinguishability for some time interval can always be found if the map is not CP-divisible. However, it is worth noting that the commonly studied examples of open quantum system dynamics do not fall into this group of evolutions. In fact, most of the commonly studied evolutions are described by dynamical maps that are bijective for all times, for which the theorem applies. In such evolutions the equilibrium state is reached only asymptotically. For the commonly studied non-bijective  evolutions, the equilibrium state is reached in a finite time but it is not a stationary state, so the system keeps evolving. An example where this occurs is the exact amplitude damping model on resonance with Lorentzian
reservoir spectrum, where the population of the excited state undergoes damped oscillations periodically reaching zero and exhibiting revivals. 
This situation corresponds to evolution described by a dynamical map that is invertible for all times except some isolated points, when the state of the system is in a ground state. 
In this case the dimension of the image of the dynamical map $\Lambda_t$ increases from $0$ to $3$ between the i-th time $s_i$ that the system reaches the ground state and later times $t$
in the interval $s_i < t < s_{i+1}$ between $s_i$ and the (i+1)-th time $s_{i+1}$ that the system reaches the ground state.

\section{Discussion}

In this work, we have demonstrated the equivalence of two a priori complementary concepts of quantum Markovianity, CP-divisibility, which provides a mathematical description of dynamical map, and information flow, which gives a physical interpretation to the memoryless feature of Markovian dynamics, for the case of bijective dynamical maps. In what follows we emphasise and discuss the advances in comparison with the results known from the literature.

As explained above, the case of information backflow for bijective dynamical maps has been considered in Ref. \cite{RChK}. However, they do not explicitly ensure that whenever the CP-divisibility of a dynamical map is violated at a time interval, there exist initial states that witness it during the actual evolution. 
This issue of observability of information backflows was successfully addressed in Ref. \cite{Buscemi&Datta} without the assumption of bijectivity of the dynamical maps.
Unfortunately, no recipe is provided to construct the required ensemble of initial states and, in particular, no upper bound on its size is given other than the dimension of the operator space $d^{4}$. This severely limits the operational consequences of the result as it makes it practically very demanding to employ. Our results do not suffer from these issues and prove that the simplest ensemble consisting of two initial states with the same prior probability suffices to determine if a bijective dynamical map is CP-divisible, even without entanglement. 

The cost of obtaining the above advantages is an increase of the dimensionality of the ancillary system from $d$ to $d+1$. However, the payoff, apart from all the above results, is that for any bijective dynamical map that is not CP-divisible, our results provide a construction of pairs of initial states that detect the information backflow. Finally, we showed that any bijective non-CP-divisible dynamical map can be witnessed by a pair of separable states. This result goes against intuition, since the CP property of dynamical maps is necessary due to possible entanglement between the evolving system and an ancilla.

\section*{Acknowledgements}
The authors are grateful to {\'A}ngel Rivas for useful correspondence.
The authors also thank Francesco Buscemi and Marco Piani for comments.
Support from the ERC CoG QITBOX, the AXA Chair in Quantum Information Science, Spanish MINECO (FOQUS FIS2013-46768-P and SEV-2015-0522), Fundacion Cellex, and the Generalitat de Catalunya (SGR 875) is acknowledged. B.B. acknowledges financial support
from  the  Polish  Ministry  of  Science  and  Higher  Education
(The "Mobility Plus" Program grant no 1107/MOB/2013/0). M.J. acknowledges support from the Marie Curie COFUND action through the ICFOnest program and the John Templeton Foundation.

\section{Appendix}

\subsection{Terminology}\label{app1}

Let $\mathcal{H}=\mathbb{C}^{d}\otimes\mathbb{C}^{d+1}$ be the Hilbert space of the system and ancilla where $\mathbb{C}^{d}$ is the Hilbert space $\mathcal{H}_{S}$ of the system and $\mathbb{C}^{d+1}$ is the Hilbert space of the ancilla.
Furthermore, let $\mathcal{H}_{A}$ and $\mathcal{H}_{A'}$ be two orthogonal subspaces isomorphic to $\mathbb{C}^d$ and  $\mathbb{C}$, respectively, of the ancilla Hilbert space $\mathbb{C}^{d+1}$. Let $\mathcal{H}_{A'}$ be spanned by the vector $|d+1\rangle$.

Let $L(\mathcal{H}_{S})$ be the space of linear operators on $\mathcal{H}_{S}$. Similarly let $L(\mathcal{H}_{S}\otimes \mathcal{H}_{A})$ and $L(\mathcal{H})$ be the spaces of linear operators on $\mathcal{H}_{A}\otimes \mathcal{H}_{S}$ and $\mathcal{H}$, respectively. The subset of positive semi-definite trace one linear operators is the set of physical states.

Let $\Lambda_s:L(\mathcal{H}_{S})\to L(\mathcal{H}_{S})$ and $\Lambda_t:L(\mathcal{H}_{S})\to L(\mathcal{H}_{S})$ be CP maps.
Furthermore, let $I(\Lambda_s)$ be the image of $\Lambda_s$ acting on $L(\mathcal{H}_{S})$.
If $\Lambda_s$ is not bijective the image is of lower dimension than $L(\mathcal{H}_{S})$ and is contained in a unique hyperplane $P[I(\Lambda_s)]$ of the same dimension as the image.
We also consider the extension $\mathcal{I}_d\otimes\Lambda_s$ of the map $\Lambda_s$ to the space $L(\mathcal{H}_{A}\otimes \mathcal{H}_{S})$.
We denote the image of $\mathcal{I}_d\otimes\Lambda_s$ acting on $L(\mathcal{H}_{A}\otimes \mathcal{H}_{S})$ by $I(\mathcal{I}_d\otimes\Lambda_s)$.
Finally, let $\mathcal{I}_{d+1}\otimes \Lambda_s$ be the extension of $\Lambda_s$ to $L(\mathcal{H})$ and let
$I(\mathcal{I}_{d+1}\otimes \Lambda_s)$ be its image.

Now we note that we can let an operator in the hyperplane $P[I(\Lambda_s)]$ be the point of origin in $L(\mathcal{H}_{S})$ and view the difference between any given operator in $L(\mathcal{H}_{S})$ and this origin as a vector. When $L(\mathcal{H}_{S})$ is viewed as a vector space in this way the hyperplane $P[I(\Lambda_s)]$ is a subspace. Given this description we can express the lowest dimensional hyperplane $P[I(\mathcal{I}_{d+1}\otimes \Lambda_s)]$ in $L(\mathcal{H})$ that contains $I(\mathcal{I}_{d+1}\otimes \Lambda_s)$ as the tensor product $L(\mathbb{C}^{d+1})\otimes P[I(\Lambda_s)]$.

\subsection{Proof of Lemma 1}
\label{sep}
The set of separable states is convex and of the same dimensionality as the full set of states. Consider an orthogonal basis $\{e_i\}$, which has the smallest possible number of basis vectors compatible with the full set of states being contained in $\SPAN(\{e_i\})$. Then the set of separable states cannot be contained in any subspace spanned by less than the full set of the basis vectors. 

Any linear map can be described by its action on the set of basis vectors. Therefore, for a linear map that maps separable states to separable states, the dimensionality of the set of separable states changes in the same ways as the dimensionality of the full set of states.
If $\Lambda_s$ is a dynamical map such that $P[I(\mathcal{I}_{d+1}\otimes \Lambda_s)]$ is the lowest dimensional hyperplane that contains $I(\mathcal{I}_{d+1}\otimes\Lambda_s)$, then $P[I(\mathcal{I}_{d+1}\otimes \Lambda_s)]$ is also the lowest dimensional hyperplane that contains the image of the set of the separable states under $\mathcal{I}_{d+1}\otimes\Lambda_s$. Moreover, the image of a convex set under a linear map is convex and $I(\mathcal{I}_{d+1}\otimes\Lambda_s)$ is always of positive dimension. Therefore, there exists a state $\sigma$ such that it is contained in a set $X\subset I(\mathcal{I}_{d+1}\otimes\Lambda_s)$ that contains only separable states and is open when considered as a subset of $P[I(\mathcal{I}_{d+1}\otimes \Lambda_s)]$.

We can now make the following observation. Assume that there exists two times $s$ and $t$, $s<t$, and two states $\rho_1$ and $\rho_2$ such that $||\mathcal{I}_{d+1}\otimes\Lambda_s(\rho_1-\rho_2)||_1<||\mathcal{I}_{d+1}\otimes\Lambda_t(\rho_1-\rho_2)||_1$.
In particular, Theorem \ref{theo} implies that if the evolution is bijective and not CP-divisible there exists two such states.
If we construct two new states $\rho'=(1-p)\sigma+p\rho_1$ and $\rho''=(1-p)\sigma+p\rho_2$ it is then clear that $||\mathcal{I}_{d+1}\otimes\Lambda_s(\rho'-\rho'')||_1<||\mathcal{I}_{d+1}\otimes\Lambda_t(\rho'-\rho'')||_1$. Moreover, for some sufficiently small $p$ both $\rho'$ and $\rho''$ are in the image of separable states under $\mathcal{I}_{d+1}\otimes\Lambda_s$ which in turn is a subset of the separable states contained in $I(\mathcal{I}_{d+1}\otimes\Lambda_s)$.

We can thus conclude that if there exist two states for which the distinguishability as measured by the trace distance increases in some timestep, there also exist two separble states for which the distinguishability increases.

\subsection{Proof of Lemma 2}
\label{app1}
Let $\Lambda_s$ and $\Lambda_t$ be the dynamical maps describing the evolution from the initial time to time $s$ and $t$ respectively. Assume that $s<t$.

If $\Lambda_s$ is invertible there exist a linear map $V_{t,s}$ such that $\Lambda_t=V_{t,s}\Lambda_s$, given by $V_{t,s}=\Lambda_t\Lambda_s^{-1}$.
If on the other hand $\Lambda_s$ is not invertible there exist two states $\rho_1$ and $\rho_2$ such that $\mathcal{I}_{d+1}\otimes\Lambda_s(\rho_1)=\mathcal{I}_{d+1}\otimes\Lambda_s(\rho_2)$. 
Then, for every operator $\sigma$ it follows that $\mathcal{I}_{d+1}\otimes\Lambda_s[\sigma + k(\rho_1-\rho_2)]=\mathcal{I}_{d+1}\otimes\Lambda_s(\sigma)$ for $k\in{\mathbb{C}}$. In other words the operators $\sigma + k(\rho_1-\rho_2)$, $k\in{\mathbb{C}}$, defines a plane in $L(\mathcal{H})$ such that all operators in the plane are mapped to the same state by $\mathcal{I}_{d+1}\otimes\Lambda_s$. Since this is true for any $\sigma$, if follows that every operator in $L(\mathcal{H})$ belongs to a plane of this type.

If a second pair of states $(\rho_1',\rho_2')$ satisfy  $\mathcal{I}_{d+1}\otimes\Lambda_s(\rho'_1)=\mathcal{I}_{d+1}\otimes\Lambda_s(\rho'_2)$ and 
$\rho'_1-\rho_1'$ is linearly independent of the difference between the states of the first pair $\rho_1-\rho_2$, there is a second independent family of planes where each plane is defined by $\sigma + k(\rho_1'-\rho_2')$, $k\in{\mathbb{C}}$, and $\mathcal{I}_{d+1}\otimes\Lambda_s[\sigma + k(\rho_1'-\rho_2')]=\mathcal{I}_{d+1}\otimes\Lambda_s(\sigma)$.

Consider all pairs $(\rho^i_1,\rho_2^i)$ of states satisfying $\mathcal{I}_{d+1}\otimes\Lambda_s(\rho^i_1)=\mathcal{I}_{d+1}\otimes\Lambda_s(\rho^i_2)$. The operators $\rho^i_1-\rho_2^i$ corresponding to all such pairs spans a subspace of $L(\mathcal{H})$. For such a subspace we can select a basis $\{e_i\}$. Any two states $\sigma$ and $\sigma+\sum_{i}k_ie_i$, $k_i\in\mathbb{C}$, are mapped to the same state by $\mathcal{I}_{d+1}\otimes\Lambda_s$. We denote the set of states that are mapped to the same state as $\sigma$ by $W_{\sigma}$.

Next, consider the following. Assume that the map $\mathcal{I}_{d+1}\otimes\Lambda_t$ does not satisfy $\mathcal{I}_{d+1}\otimes\Lambda_t(\rho^i_1)=\mathcal{I}_{d+1}\otimes\Lambda_t(\rho^i_2)$ for one pair of states such that $\mathcal{I}_{d+1}\otimes\Lambda_s(\rho^i_1)=\mathcal{I}_{d+1}\otimes\Lambda_s(\rho^i_2)$. It follows that $||\mathcal{I}_{d+1}\otimes\Lambda_t(\rho^i_1-\rho^i_2)||_1>0$. But since $||\mathcal{I}_{d+1}\otimes\Lambda_s(\rho^i_1-\rho^i_2)||_1=0$, this implies that there is an increase in distinguishability between $s$ and $t$. Note that in this case there is no map $V_{t,s}$ since the evolution is one-to-many.

Now let us assume that the evolution does not increase the distinguishability.
If we assume non-increase of distinguishability it cannot be that $\mathcal{I}_{d+1}\otimes\Lambda_t(\rho^i_1)\neq \mathcal{I}_{d+1}\otimes\Lambda_t(\rho^i_2)$ while $\mathcal{I}_{d+1}\otimes\Lambda_s(\rho^i_1)=\mathcal{I}_{d+1}\otimes\Lambda_s(\rho^i_2)$ for any pair of states by the above argument. 
This means that for every $e_i$ not only $\mathcal{I}_{d+1}\otimes\Lambda_s(e_i)=0$, but also $\mathcal{I}_{d+1}\otimes\Lambda_t(e_i)=0$. Thus, every set $W_{\sigma}$ of operators that $\mathcal{I}_{d+1}\otimes\Lambda_s$ maps into a single state is also mapped to a single state by $\mathcal{I}_{d+1}\otimes\Lambda_t$.

This property allows us to construct a map from a state $\mathcal{I}_{d+1}\otimes\Lambda_s({\sigma})$ to the set $W_{\sigma}$, and from $W_{\sigma}$ to the state $\mathcal{I}_{d+1}\otimes\Lambda_t({\sigma})$. That is, we have a map from the states $\mathcal{I}_{d+1}\otimes\Lambda_s({\sigma})$ to the states $\mathcal{I}_{d+1}\otimes\Lambda_t({\sigma})$.

Therefore, we can now replace the domain of $\mathcal{I}_{d+1}\otimes\Lambda_s$ and $\mathcal{I}_{d+1}\otimes\Lambda_t$ by a domain where each set
$W_{\sigma}$ is replaced by a single operator. For example, for each $e_i$ we can make a projection onto the orthogonal complement of $e_i$. Let $\mathcal{I}_{d+1}\otimes\tilde{\Lambda}_s$ and $\mathcal{I}_{d+1}\otimes\tilde{\Lambda}_t$ denote the linear maps on the new domain.
Now we note that $\mathcal{I}_{d+1}\otimes\tilde{\Lambda}_s$, and therefore $\tilde{\Lambda}_s$, is an invertible map, and thus the map $V_{t,s}$ is given by $V_{t,s}=\tilde{\Lambda}_t\tilde{\Lambda}_s^{-1}$.

We can therefore conclude that when the evolution is such that the distinguishability is non-increasing there exist a linear map $V_{t,s}$. If the evolution is such that the distinguishability increases, the map $V_{t,s}$ may or may not exist.

\subsection{Addendum to the Theorem of Reference \cite{RChK} }
\label{app3}

Assume that $||{\mathcal{I}_d}\otimes V_{t,s}(\Delta_{p})||_1>||\Delta_p||_1$ for the Helstrom matrix $\Delta_p=p\rho_1-(1-p)\rho_2$.
If $\rho_1$ and $\rho_2$ are not in $I({\mathcal{I}_d}\otimes\Lambda_{s})$ we can consider the following construction. We rewrite $\Delta$ as $p\rho_1-(1-p)\rho_2=(1/r-1)p\sigma+p\rho_1-(1/r-1)p\sigma-(1-p)\rho_2$ where $\sigma$ is a state in the interior of $I({\mathcal{I}_d}\otimes\Lambda_{s})$. Then, we multiply by a constant $r/(2p+r-2rp)=1/Tr[(1/r-1)p\sigma+p\rho_1+(1/r-1)p\sigma+(1-p)\rho_2]$ and re-express it as

\begin{eqnarray}
\frac{r}{2p+r-2rp}\Delta_p=\frac{1}{2p+r-2rp}[(1-r)p\sigma+rp\rho_1]\nonumber\\
-\frac{1}{2p+r-2rp}[(1-r)p\sigma+r(1-p)\rho_2]=\nonumber\\
y[(1-r)\sigma+r\rho_1]-(1-y)[(1-x)\sigma+x\rho_2],
\end{eqnarray}
where $x=r(1-p)/(p+r-2rp)$ and $y=p/(2p+r-2rp)$. The constant $r/(2p+r-2rp)$ was chosen such that for $0\leq {r}\leq{1}$ both $(1-r)\sigma+r\rho_1$ and $(1-x)\sigma+x\rho_2$ are positive hermitian trace one operators.

Now, we observe that $x$ goes to zero continuously when $r$ goes to zero continuously. Therefore both $(1-r)\sigma+r\rho_1$ and $(1-x)\sigma+x\rho_2$ goes towards $\sigma$ when $r$ goes towards zero. 

There is therefore some sufficiently small $r$ for which $(1-r)\sigma+r\rho_1$ and $(1-x)\sigma+x\rho_2$ are both inside $I({\mathcal{I}_d}\otimes\Lambda_{s})$.
For such an $r$ we can thus interpret $r/(2p+r-2rp)\Delta$ as a Helstrom matrix for two operators in $I({\mathcal{I}_d}\otimes\Lambda_{s})$.

Thus, the existence of a Helstrom matrix $\Delta_p=p\rho_1-(1-p)\rho_2$ such that $||{\mathcal{I}_d}\otimes V_{t,s}(\Delta_p)||_1>||\Delta_p||_1$ implies the existence of another Helstrom matrix $\Delta'_{y}=y\rho_1'-(1-y)\rho_2'$ where $\rho_1'$ and $\rho_2'$ are in $I({\mathcal{I}_d}\otimes\Lambda_{s})$ and $||{\mathcal{I}_d}\otimes V_{t,s}(\Delta'_y)||_1>||\Delta'_y||_1$.

\end{document}